\documentclass[aps,prl,twocolumn,superscriptaddress,showpacs]{revtex4-1}

\usepackage{graphicx}
\usepackage{dcolumn}
\usepackage{bm}
\usepackage{float}
\usepackage{amssymb}
\usepackage{amsmath}
\usepackage{hyperref}
\usepackage[mathlines]{lineno}

\begin{document}

\title{Measurement of the Dynamical Structure Factor of a 1D Interacting Fermi Gas}




\author{T. L. Yang}
\affiliation{Department of Physics and Astronomy, Rice University, Houston, Texas 77005, USA}

\author{P. Gri\v{s}ins}
\affiliation{Department of Quantum Matter Physics, University of Geneva, 1211 Gen\`{e}ve, Switzerland}

\author{Y. T. Chang}
\affiliation{Department of Physics and Astronomy, Rice University, Houston, Texas 77005, USA}

\author{Z. H. Zhao}
\affiliation{Department of Physics and Astronomy, Rice University, Houston, Texas 77005, USA}

\author{C. Y. Shih}
\affiliation{Department of Physics and Astronomy, Rice University, Houston, Texas 77005, USA}
\altaffiliation{Institute for Quantum Computing, University of Waterloo, West Waterloo, Ontario, Canada}

\author{T. Giamarchi}
\affiliation{Department of Quantum Matter Physics, University of Geneva, 1211 Gen\`{e}ve, Switzerland}

\author{R. G. Hulet}
\affiliation{Department of Physics and Astronomy, Rice University, Houston, Texas 77005, USA}
\email{randy@rice.edu}

\date{\today}
\begin{abstract}
We present measurements of the dynamical structure factor $S(q,\omega)$ of an
interacting one-dimensional (1D) Fermi gas for small excitation energies. We use the two lowest
hyperfine levels of the $^6$Li atom to form a pseudo-spin-1/2 system whose $s$-wave
interactions are tunable via a Feshbach resonance. The atoms are confined to 1D by a
two-dimensional optical lattice. Bragg spectroscopy is used to measure a response of the gas to density
(``charge'') mode excitations at a momentum $q$ and frequency $\omega$, as a function of the interaction strength.
The spectrum is obtained by varying $\omega$, while the angle between two laser beams determines $q$,
which is fixed to be less than the Fermi momentum $k_\textrm{F}$. The measurements agree well with Tomonaga-Luttinger theory.

\end{abstract}
\maketitle
Understanding many-body systems is one of the most important challenges of quantum physics, not only intellectually, but also on how it impacts our ability to develop materials with novel properties. In high dimensions, the cornerstone of our understanding of such systems has been the existence of excitations behaving much like the original individual free particles of which it is comprised. For fermions these are the Landau quasiparticles of Landau's Fermi liquid theory \cite{landau_fermiliquid_theory_static,nozieres_book}. In this case, interactions mostly result in the modification of parameters, such as mass.

Very different behavior occurs when the dimensionality of the system is reduced, which reinforces the effects of interactions.  In one-dimension, even the physics of nearly free particles may be substantially altered \cite{Giamarchi_Book}. Because all particles are affected by interactions, in reduced dimensions excitations of the system are collective, while no individual, single particle-like excitations can occur. This leads to a type of physics, known as the Tomonaga-Luttinger liquid (TLL) \cite{Tomonaga1950,Luttinger1963,haldane_bosonisation}, in which all the excitations are decomposed into collective excitations of charge and spin. As a consequence, the system is critical with correlations decreasing as a power law at zero temperature, and individual excitations fractionalize, decomposing into products of topological excitations carrying spin but no charge (spinons) and charge but no spin (holons) \cite{Giamarchi_Book}. The physics of such systems is thus controlled by the velocities of the charge and spin collective excitations as well as a dimensionless parameter controlling the power law decay of the correlations. These parameters are directly related to the microscopic interaction in the system \cite{haldane_exponents_spin_chain,Giamarchi_Book}.

Observing these properties presents considerable challenges. In condensed matter, the power law behaviors attributed to TLL have been seen, in particular, in organic conductors \cite{schwartz_electrodynamics}, nanotubes \cite{yao_nanotube_kink}, and in the conductivity of edge states \cite{milliken_edge_states}, while spin-charge separation was observed in quantum wires \cite{Auslaender2005}. Further examples of experimental realizations of TLL can be found in \cite{giamarchi_cras_1d}. Due to the screened nature of the long-range Coulomb interactions in electronic systems, however, it is difficult to make quantitative comparisons with the TLL theory. In particular, control of interactions has not been previously accomplished. It is thus important to have experimental systems in which such quantitative comparisons can be made. In condensed matter, quantum spin systems provide an excellent route to quantitatively test for TLL \cite{bouillot_dynamical_ladder,povarov_temporal_scaling_DIMPY} but only for interacting bosonic-like systems.

Given their remarkable degree of control of lattice structure and interactions \cite{bloch_cold_lattice_review}, cold atoms provide a promising, complementary route to tackle TLL physics. The contact nature of interactions in cold atoms, and their control via a Feshbach resonance \cite{Chin2010} enables detailed comparison between the theoretical predictions of TLL and experiment.  The possibility of measuring the dynamical structure factors of charge and spin provides a check that collective excitations indeed control the entire excitation spectrum. Realizing these measurements, however, has not proven to be easy. For bosonic systems, quasi-long range behavior of correlation functions consistent with TLL predictions have been observed for optical lattices \cite{paredes04_tonks_gas} and for atom chips \cite{hofferberth_interferences_atomchip_LL}, and the sound velocity connected to TLL physics \cite{bing_sound_TLL_bosons_cold} and the dynamical structure factor \cite{Fabbri2015} were recently measured.  For fermionic systems, the dynamical structure factor was probed for SU(N) fermions~\cite{Pagano2014}, but due to the absence of a Feshbach resonance, a detailed analysis of the excitation spectrum as a function of the interaction strength has been missing.


In this paper, we use Bragg spectroscopy \cite{Stenger1999,Brunello2001} to measure the
dynamical structure factor $S(q,\omega)$ of the density (``charge'') mode of a one-dimensional
(1D) Fermi gas of fermionic $^6$Li whose repulsive $s$-wave interactions are broadly tunable.
Bragg spectroscopy, which employs two-photon stimulated transitions, is a sensitive method to
detect density fluctuations, and thus $S(q,\omega)$ in cold atom systems. Our measurement of
$S(q,\omega)$ for $q < k_\textrm{F}$, where $k_\textrm{F}$ is the Fermi momentum, agrees
well with the TLL theory when using the local density approximation to account for the density
inhomogeneity of the trapped gas, thus providing the first exploration of such physics with cold atomic fermions with tunable interactions.

The 1D experiment of Ref.~\cite{Pagano2014} employed Bragg spectroscopy to
measure the low-energy energy excitation spectra of the density mode of a 1D Fermi gas with a
tunable number of spin components. Bragg spectroscopy has also been used to measure the
dynamical structure factor in 3D for both the spin and density excitations of a strongly
interacting Fermi gas at high momentum, where $q \gg k_\textrm{F}$, \cite{Hoinka2012}, and more
recently, the charge excitation spectrum for $q \lesssim k_\textrm{F}$ \cite{Hoinka2017}.

The two lowest hyperfine sub-levels ($|F=1, m_F=1/2\rangle$ and $|F=1, m_F=-1/2\rangle$) of $^6$Li
constitute a quasi-stable, pseudo-spin-1/2 system, which we label as $|\nobreak\uparrow\nobreak\rangle$ and
$|\nobreak\downarrow\nobreak\rangle$, respectively. The three-dimensional $s$-wave scattering length, $a$, for
states of opposite spin projection can be tuned from $a=0$ at 528 G, to $a \simeq 400$ $a_0$ at
606 G, where $a_0$ is the Bohr radius \cite{Zurn13}. Beyond this repulsive interaction
strength, however, the ``upper branch'' of the Feshbach resonance becomes unstable to the
formation of deeply bound dimers \cite{Hart15}.

The apparatus and some of the methods used in this experiment have been described previously
\cite{Hart15,Duarte15}. We focus here on the primary differences. The atoms are loaded into a
crossed beam optical trap formed from three mutually orthogonal infrared (IR) laser beams.
These beams are retroreflected, but the polarization of each retroreflected beam is rotated by
90$^{\circ}$ to form a trap without a lattice. After loading this trap, we measure the total number
of atoms and their temperature to be $1.4 \times 10^5$ and 0.05 $T_\textrm{F}$, respectively,
where $T_\textrm{F}$ is the Fermi temperature of each spin state assuming no interactions. We then increase the depth of
the trap and rotate the polarization of the retroreflected beams to form a 3D optical lattice with a
lattice depth $V_\textrm{L} = 7 E_\textrm{r}$, where $E_\textrm{r} = h^2/(2m\lambda^2)$ is
the recoil energy, $h$ is Planck's constant, $m$ is the atomic mass, and $\lambda = 1,064$ nm is
the wavelength of the light.  During this process, we also adjust the scattering length to desired
final value before the lattice depth reaches $V_\textrm{L} = 2.5 E_\textrm{r}$.
We superimpose a blue-detuned (532 nm), non-retroreflected laser
beam along each axis to partially compensate the overall confining envelope of the infrared
beams \cite{Mathy2012,Hart15}. The intensities of the compensating beams are adjusted to flatten the
confining potential and to tune the central density in the lattice to be near one atom per site, $n
\simeq 1$, for each interaction strength. Previous measurements showed that the atoms formed a
Mott insulator for $a \gtrsim 250 a_\textrm{o}$ at $V_\textrm{L} = 7 E_\textrm{r}$ \cite{Duarte15}.
The small variation of density for different interactions is an important feature of our experiment.

We then slowly turn off the compensating green beams and the vertical IR beam while
simultaneously increasing the intensity of the two remaining lattice beams to form a 2D lattice
with $V_\textrm{L} = 15 E_\textrm{r}$. The 2D lattice creates a bundle of nearly isolated 1D tubes,
characterized by axial and radial harmonic frequencies of $\omega_\textrm{z} = (2\pi) 1.3$
kHz, and $\omega_\perp = (2\pi) 198$ kHz, respectively. The final total atom number is
$N=1.1 \times 10^5$. The total number of atoms in the central tube, $N_m$, is nearly independent
of $a$ for stronger interactions, but is somewhat larger for smaller $a$.
By performing an inverse Abel transform on column density images we
determine the 3D density distributions, and find that $N_m$ varies from 55, at larger $a$, to 70
for an ideal gas. The most probable atom number per tube is between 37 and 40 for all interactions.
Taking $N_m = 60$ gives $k_B T_F = 1.9~\mu $K for each spin-state, which is much less than $\hbar\omega_\perp = 9.5~\mu $K.

\begin{figure}
\includegraphics[width=0.48\textwidth]{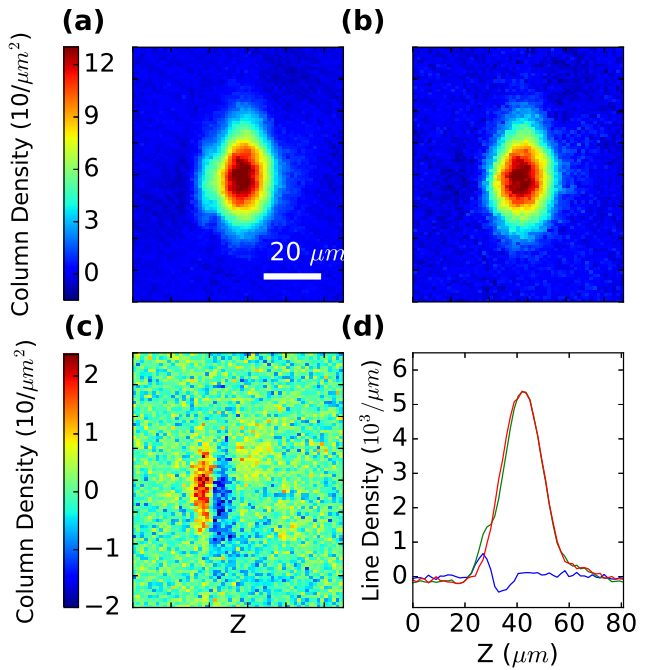}
\caption{Column density images of (a) signal and (b) reference, corresponding to Bragg, or no Bragg pulse, respectively.  (c) Difference of (a) and (b). (d) Line density of reference (red), signal (green), and their difference (blue). The line densities are calculated by summing over the column densities along the axis perpendicular to the 1D tube direction, which is the vertical axis for (a), (b), and (c). The Bragg signal is proportional to the area under the positive portion of the difference line density curve.}
\label{fig:images}
\end{figure}

\begin{figure} \includegraphics[width=0.48\textwidth]{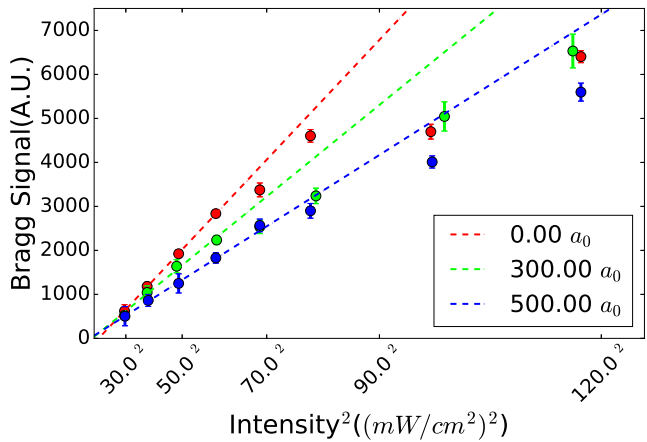}
\caption{\label{fig:intensity}  Bragg signal vs. intensity per Bragg beam. The plotted signal is an average of the signal at three different frequencies: $\omega/2\pi =$~5, 9, and 13 kHz. The dashed lines are linear fits for laser intensities below 65 mW/cm$^2$. The two Bragg beams are Gaussian, each with a waist of 570 $\mu$m, and the pulse time is 300 $\mu$s.} \end{figure}

\begin{figure} \includegraphics[width=0.48\textwidth]{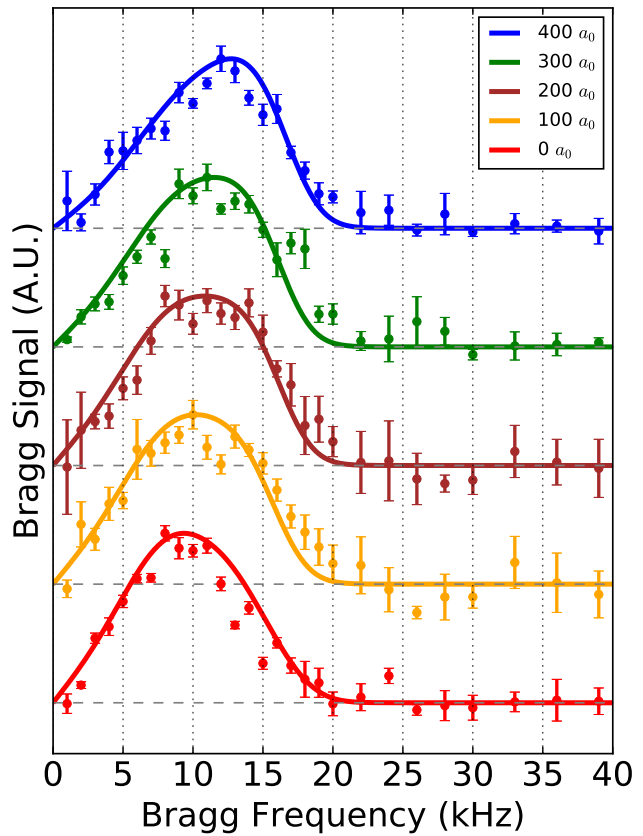}
\caption{\label{fig:spectra} (color online). Bragg spectra vs. $\omega/2\pi$. The Bragg signal is proportional to the momentum transfer. Each data point is an average of 20-30 experimental shots. The error bars are calculated from the bootstrapping method~\cite{Efron1979}. The theoretical spectra (solid lines) are a result of using the LDA with TLL theory for each interaction, and a temperature of 200 nK. There are no additional fitting parameters other than overall scaling.}  \end{figure}

\begin{figure} \includegraphics[width=0.48\textwidth]{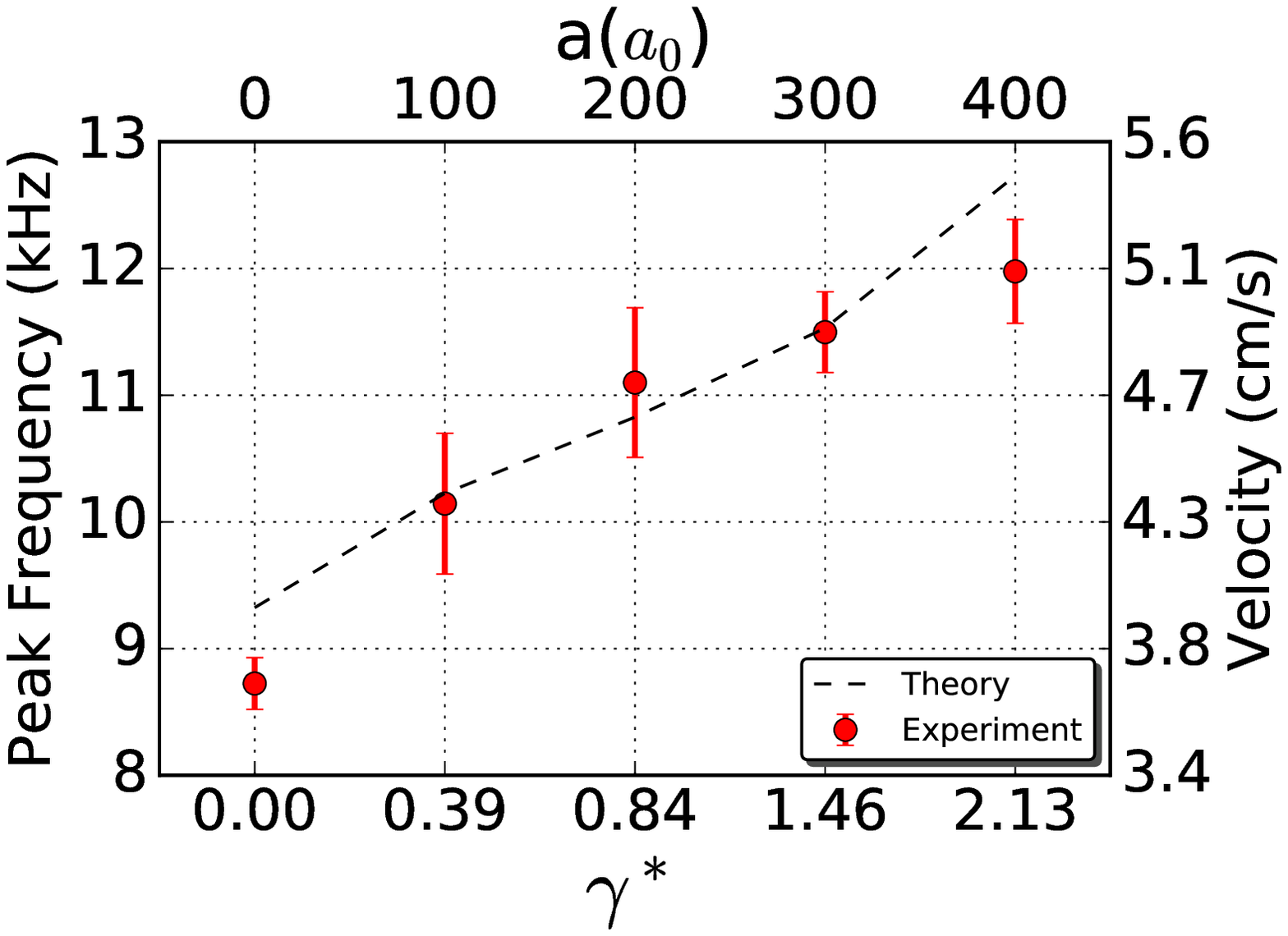}
\caption{(color online). Peak frequency of each spectrum derived from empirical fits of the measured excitation spectra from Fig. ~\ref{fig:spectra}. For the theory curve (dashed), we simply find the location of the maximum excitation. $\gamma^*$ corresponds to the dimensionless interaction parameter $\gamma$ at the center of the 1D tube with the most probable atom number. Here,
$\gamma = \frac{m g_1(a)}{\hbar^2 \rho_{1D}}$ and $g_1 = \frac{2\hbar\omega_\perp a}{1-1.03a/a_\perp}$ where $a_{\perp} = \sqrt{\frac{\hbar}{m\omega_\perp}}$ and $\rho_{1D}$ is the total 1D density \cite{Dunjko2001}. The corresponding speed of sound, $\omega/q$, is given by the right axis.}
\label{fig:most_probable}
\end{figure}
Bragg spectroscopy employs two laser beams, with wavevectors $\mathbf{k_1}$ and $\mathbf{k_
2}$ and frequency difference $\omega$. They propagate with an angle $\theta$ between them and
intersect the atoms symmetrically about a line perpendicular to the tube ($z$) axis, and thus possess a net momentum along the z-direction. These beams
drive a stimulated two-photon transition that couples the ground state to an excitation of
frequency $\omega$ and z-component of momentum $q = |\mathbf{k_1} - \mathbf{k_2}| = 2k
\sin(\theta/2)$, where $k = |\mathbf{k_1}| = |\mathbf{k_2}|$. We fix the angle $\theta/2 \simeq
4.5^\circ$ such that $q/k_\textrm{F} \simeq 0.2$ for a central tube filling of $N_m = 60$.

The Bragg beams are detuned by 14 GHz blue of the D$_2$ transition in $^6$Li in order to
minimize the rate of spontaneous emission. Since the detuning is large compared with the 76
MHz splitting between the $|\uparrow\rangle$ and $|\downarrow\rangle$ states, the signal is
sensitive only to density excitations and not to spin. The Bragg beams are pulsed on for 300
$\mu$s, which is less than half the axial period, but long compared with $\omega^{-1}$,
conditions that both simplify the analysis \cite{Brunello2001} and minimize the pulse-time
broadening. The optical lattice beams are switched off immediately following the Bragg pulse.
After 150 $\mu$s of time of flight, we image the column density distribution of both states using
phase-contrast polarization imaging \cite{Bradley1997} with a probe detuning of 33 MHz red of
the $|\downarrow\rangle$ imaging transition (and 110 MHz red of the $|\uparrow\rangle$ transition).
By repeating the experiment without Bragg beams,
we record a reference image that we subtract from the signal image obtained with the Bragg
beams present, as shown in Fig. ~\ref{fig:images}. These column density images are integrated
in the direction transverse to the tube axis to obtain axial line densities shown in
Fig.~\ref{fig:images}(d). The integral under the positive part of the line densities are proportional to
the total momentum transferred by the Bragg beams, and is what we define as ``signal''.

We checked that the Bragg signal is in the linear response regime by varying the Bragg beam intensity,
while fixing the pulse duration. In this regime, the rate of stimulated Bragg transitions should depend
quadratically on the laser intensity (assumed equal for each beam). As shown in
Fig.~\ref{fig:intensity}, an intensity per beam of less than 55 mW/cm$^2$ ensures that the
momentum transfer is in the linear response regime over the entire range of interaction strengths
accessed in the experiment. We fix the intensity at this value.

The Bragg spectra for five values of the interaction parameter are presented in Fig.~\ref{fig:spectra}. Each data point corresponds to an average of 20-30 experimental runs for
each value of $\omega$ and fixed $q$. The full-width at half maximum of these spectra range
from 9 kHz for a noninteracting gas to 11 kHz for the most strongly interacting one. These
spectral widths are large compared to the pulse-time broadening of 3 kHz. The spectra are empirically found to fit well to a skew normal distribution (convolution of a Gaussian with the error function). The most probable value of  $\omega$ may be obtained from the fitting parameters for each interaction, and these are plotted in Fig. ~\ref{fig:most_probable}. The most probable
frequency increases with interaction strength until $a = 400 a_0$. We notice heating
and atom loss beyond this interaction, quite probably due to three-body recombination from the
unstable upper branch during the transition from the 3D to 2D lattice. In contrast,
we observe no atom loss for $a$ between 0 and 400 $a_0$.

In the linear response regime, the experimentally measured momentum transfer for each $q$
and $\omega$ is proportional to $S(q,\omega) - S(-q,-\omega)$, where the second term accounts for
inverse Bragg scattering: absorption by beam 2 and stimulated emission by beam 1
\cite{Brunello2001}. The second term will be small compared to the first when $k_B T \ll
\hbar\omega$ \cite{Bobrov1990,Brunello2001}, and in this case, the measured Bragg signal is proportional to
$S(q,\omega)$.

In order to compare with theory, an inverse Abel transform is applied to the
measured column densities (without a Bragg pulse) to obtain the distribution of atom number per
tube at each interaction strength. Using the local density approximation (LDA) we split each tube into about a hundred pieces of length $dz$ with constant density $n(z)$,
calculate the dynamic structure factors independently for each piece, and sum them together to get the momentum transfer for one tube.
As momentum is an additive quantity, we repeat this procedure for each tube in the lattice,
summing the individual momentum transfers into the resulting total momentum transfer.

For a homogeneous Fermi gas, the dynamic structure factor is:
\begin{equation}
S(q,\omega, k_F, T, N) = \dfrac{\mathrm{Im}\ \chi(q, \omega, k_F, T, N)}{\pi(1 - e^{-\beta\hbar\omega})},
\label{eq:s}
\end{equation}
where $\chi$ is the dynamic susceptibility~\cite{Cherny2006}. The TLL theory \cite{Giamarchi_Book} states that at small $q$ the susceptibility is dominated by a collective charge mode, whose velocity has a very precise interaction dependence that can be computed exactly with known interactions~\cite{Guan2013}.  For a homogeneous system at zero temperature and small $q$, the susceptibility has a resonance at $\omega = u q$ giving direct and convenient access to the velocity of charge excitations.  Thus, for weak interactions, Eq. \ref{eq:s} may be used to calculate the structure factor, but in this case, substituting the speed of sound $u$ obtained from Bethe Ansatz~\cite{Guan2013} for the Fermi velocity to account for the shift in the resonance.  The velocity $u$ may be calculated exactly as a function of interaction.  More details of our theoretical analysis are available in the Supplemental Materials~\footnote{See Supplemental Material [URL] for additional details of our theoretical analysis, which includes Ref.~\cite{Caux2006}.}.

We use this procedure to calculate the structure factor for a temperature of 200 nK, and compare with the experimental data, as shown in Fig. \ref{fig:spectra}.  The only adjustable parameter is the overall scaling of the excitation.  The agreement between the calculated lineshape with the experimental one validates our method to account for the sources of broadening, and shows that the experiment gives direct access to the interaction dependence of the velocity.  The peaks of the experimental excitation spectra for each interaction are plotted in Fig. \ref{fig:most_probable} together with the theoretical result. The agreement between the measured and computed interaction dependence of the velocity is very good and provides the first experimental test of the change in velocity of the collective excitation of a 1D Fermi gas vs. interactions.


We also attempted to measure the dynamical structure factor of the spin mode by adjusting the detuning of the Bragg laser to be negative for one spin state, while positive for the other \cite{Hart15}.  Since the two optical transitions are separated by only 76 MHz, however, we were unable to observe a Bragg signal without destroying the sample with excessive spontaneous emission from the excited 2P state.  It may be possible to observe a spin-dependent Bragg signal in the future by detuning from the 3P excited state instead, as the rate of spontaneous emission is reduced by the ratio of linewidths, which is a factor of 8 in this case \cite{Duarte2011}. Such a measurement will thus give full access to the two collective modes controlling the physics of the interacting fermionic system. Furthermore, it would be interesting to couple such measurements with those of the single particle excitation
spectrum, e.g. by momentum resolved RF spectroscopy~\cite{Frohlich2012,Gaebler2010}, to establish the link between the single particle spectrum and the collective modes predicted by TLL.

In conclusion, we have measured the dynamic response of a one-dimensional two-component fermionic system using Bragg spectroscopy and find good agreement with TLL theory for the collective charge mode.
The ability to adjust the interaction strength via a Feshbach resonance enables future studies such as a direct observation of spin-charge separation, the dynamic response for high q excitation that goes beyond the Luttinger liquid theory, or possibly a system with p-wave interactions for a single spin state.

\begin{acknowledgments}

This work was supported in part by the Army Research Office Multidisciplinary University Research Initiative (Grant No. W911NF-14-1-0003), the Office of Naval Research, the NSF (Grant No. PHY-1707992), and the Swiss National Science Foundation under division II.

\end{acknowledgments}

\bibliography{1DBragg}

\end{document}


\begin{center}
\textbf{\large Supplemental Materials}
\end{center}
\section*{Tomanaga Luttinger Liquid Theory}

In order to precisely test for the interaction dependence of the velocity and relate the measurements to the TLL theory, we need to take into account the broadening of the resonance in the susceptibility.

The TLL theory would lead to a structure factor which has a strong resonance (a $\delta$-function peak in the asymptotic limit of the TLL low energy approximation) at a frequency of the form $\omega = v_\rho q$ where $q$ is the momentum and $v_\rho$ the velocity of the charge excitation (the holon) \cite{giamarchi_book_1d}. The renormalized velocity $v_\rho$ itself can be extracted \emph{exactly} from the Bethe-ansatz solution of the two component fermionic model in the continuum (with a fixed and constant density) since such a model is integrable \cite{guan_review_bethe}. If the model was at zero temperature and fully homogeneous, we could thus get the position of the resonance exactly from the knowledge of the interaction strength and the density.

Due to effects going beyond the TLL theory this resonance is broadened, so to compute the lineshape and be able to make a \emph{quantitative} comparison with the experiments we also need to take into account this broadening.
The broadening has three main sources: i) the intrinsic broadening of the resonance that occurs at finite $q$ and which comes from terms beyond the TLL theory (effects of band curvature etc.);  ii) the temperature broadening (which is in principle, of course, coupled to the previous source of broadening) but due to the (relatively) high temperature in the experiment is quite important; iii) the (parasitic) experimental broadening produced by inhomogeneities due to the trap.

The last source of broadening is the easiest to take into account. As discussed in the text, we take inhomogeneities into account by using an LDA approximation (both inside a single tube as a function of the position in the tube with respect to the axial trapping potential, and summing over different tubes as a function of the radial distance to the center of the trap. Such an approximation is quite standard and is expected to be quite accurate if the confining potential is smoothly varying compared to the intrinsic scales of the fermionic system, which is the case in this experiment.

In order to take into account i) and ii) we would in principle need to compute the lineshape (at a fixed value of $q$) for an interacting two component fermionic gas at finite temperature (with a fixed density).  Unfortunately, an exact computation of such broadening for two component fermionic systems, even if the system is integrable, is a considerable challenge. Even for the simpler case of bosons (Lieb-Lininger model) this calculation from Bethe-ansatz has only been achieved at zero temperature so far \cite{caux_bose_formfactors06}.

In order to take into account i) and ii) with a good degree of accuracy we use the fact that for the experiment: a) the interactions are large but not gigantic compared to the kinetic energy (as shown in Fig.~4 of the main text the dimensionless parameters $\gamma$ for the interaction is going up to essentially $\gamma = 2$); b) the main source of broadening is coming from the thermal effects and the inhomogeneities so an extremely accurate form of the intrinsic ($T=0$, homogeneous) lineshape will be largely washed out by these two additional effects. c) for a non-interacting system ($\gamma = 0$) we can compute the structure factor exactly and thus, the lineshape, since this is just the density-density correlation of free fermions (Lindhard function).

We thus assume that for the (moderate) interactions of the experiment the \emph{intrinsic} lineshape is nearly unchanged compared to the case of free fermions. The main effect of the interactions is to move the position of the resonance, which for free fermions is at $\omega = v_F q$, where $v_F$ is the Fermi velocity, to $\omega = v_\rho q$ where $v_\rho$ is the \emph{exact} velocity of the excitation depending on the interactions obtained from Bethe ansatz \cite{guan_review_bethe}. Given the importance of thermal broadening ii) and the LDA broadening iii) the error made in neglecting the interaction influence in the \emph{intrinsic lineshape} compared to the free fermion case is negligible.

We thus
\begin{itemize}
\item Compute the \emph{exact} velocity $v_\rho$ from the Bethe-ansatz solution of a problem with a fixed density $\rho_0$ and interaction $\gamma$.
\item Use this velocity instead of $v_F$ in the calculation of a free fermion structure factor at finite temperature $T$ and fixed density $\rho_0$ to get the homogeneous lineshape broadened by the effects i) and ii) at the correct position for the resonance as a function of the interactions.
\item Use the above as the input for the LDA calculation in which the density $\rho_0$ is now varied according the position inside a tube and from tube to tube as a function to the radial distance to the center of the trap according to the LDA approximation.
\item These various contributions are summed with the proper weight to obtain the full response (see Fig.~3 of the main text) that can be directly compared to the experiment.
\end{itemize}

As shown in Fig.~3 of the main text, this provides, without any adjustable parameter an excellent quantitative comparison with the lineshape observed in the experiment. As a consequence, the important parameter, namely the position of the resonance and its dependence on the interactions can be reliably extracted as shown in Fig.~4 of the main text.